%% file: main.tex
\documentclass[journal]{IEEEtran}
\usepackage{cite}

\ifCLASSINFOpdf
   \usepackage[pdftex]{graphicx}
   \DeclareGraphicsExtensions{.pdf,.jpeg,.png}
\else
\fi
\usepackage{amsmath}
\usepackage{amssymb}
\usepackage{booktabs}

\DeclareMathOperator*{\argmax}{arg\,max}
\usepackage[ruled]{algorithm2e}
\SetKwInOut{KwInput}{Input}
\SetKwInOut{KwOutput}{Output}
\usepackage{flushend}

\usepackage{array}
\usepackage{fixltx2e}

\usepackage{stfloats}
\usepackage{url}
\usepackage{csquotes}

\begin{document}
%
\title{Unsupervised active speaker detection in media content using cross-modal information}
%
%
%

\author{Rahul~Sharma,
        and~Shrikanth Narayanan,~\IEEEmembership{Fellow,~IEEE}
        \thanks{Rahul Sharma (raul.sharma@usc.edu) and Shrikanth Narayanan (shri@ee.usc.edu) is with Electrical and Computer Engineering, University of Southern California, USA}
        }

%
%

\markboth{IEEE transactions on Image Processing}%
{Shell \MakeLowercase{\textit{Sharma et al.}}: Cross modal active speaker localization}
%



\maketitle

\begin{abstract}
We present a cross-modal unsupervised framework for active speaker detection in media content such as TV shows and movies. Machine learning advances have enabled impressive performance in identifying individuals from speech and facial images. We leverage speaker identity information from speech and faces, and formulate active speaker detection as a speech-face assignment task such that the active speaker's face and the underlying speech identify the same person (character). We express the speech segments in terms of their associated speaker identity distances, from all other speech segments, to capture a relative identity structure for the video. Then we assign an active speaker's face to each speech segment from the concurrently appearing faces such that the obtained set of active speaker faces displays a similar relative identity structure. Furthermore, we propose a simple and effective approach to address speech segments where speakers are present off-screen. We evaluate the proposed system on three benchmark datasets-- Visual Person Clustering dataset, AVA-active speaker dataset, and Columbia dataset-- consisting of videos from entertainment and broadcast media, and show competitive performance to state-of-the-art fully supervised methods. 
\end{abstract}

\begin{IEEEkeywords}
unsupervised, cross-modal, speaker identity, face recognition, active speaker detection, entertainment media
\end{IEEEkeywords}

%
\IEEEpeerreviewmaketitle

\input{intro}
\input{related_work}

\input{methodology}
\input{experiments}

\input{conclusion}



%





\newpage
\ifCLASSOPTIONcaptionsoff
  \newpage
\fi



\bibliographystyle{IEEEtran}
\bibliography{ref}
\end{document}

%% file: intro.tex
\section{Introduction}
\label{sec:intro}
\IEEEPARstart{S}{peech-}related activity in a video manifests spatio-temporally both in audio and visual modalities. Mapping speech activity in the temporal domain related to \emph{when}, in time, someone is speaking is often approached from the audio modality. On the other hand, in the spatial domain, determining \emph{where} in a frame the speaker is present relies on the visual signal. In this work, we use the speech activity information obtained from the audio modality to find the corresponding speech activity in the visual modality. We address the problem of finding the speaking face (if any) from the set of candidate faces appearing in the video. We refer to this task as \emph{active speaker detection} (ASD).

ASD plays a crucial role in computational media intelligence (CMI)~\cite{somandepalli2021computational}, enabling the study of representation and portrayals of characters portrayed in media and their societal impact. Specifically, ASD is a crucial component of tools to discern \emph{who is speaking when?} in a video, such as for primary character detection, background character detection~\cite{sharma2022audio}, speaker diarization~\cite{bredin2016improving, chung2020spot, sharma2022using}, and other human-computer-interaction applications. This work focuses on ASD in entertainment media content, notably Hollywood movies and TV shows.

ASD in entertainment media content is a challenging task due to the unknown and varying number of speakers and faces in the visual frames. The speaking faces appear in diverse poses because of the rich dynamics in character interactions. The face detector fails when the speakers appear in extreme poses including looking away from the camera or moving out of the frame.
Non-ideal conditions in the audio modality, such as background (including off-screen) speakers and noisy speech, also contribute to problem's complexity.

The inherent multimodal nature of speech activity has motivated numerous approaches for active speaker detection to use the connections between the visual and audio modality. 
The idea of localizing the visual actions in the video frames responsible for the audio activity, notably speech, dominates these approaches. Audio-visual speech modeling typically involves capturing the activity in the lip region of the detected faces in the visual frames, explicitly~\cite{bendris2010lip, bredin2016improving} or implicitly~\cite{owens2018audio, zhao2018sound}, and studying its correlation with the speech waveform to discern the active speaker. Most of these approaches share the same drawbacks: i) These methods employ computationally expensive 3D Convolutional neural networks and Recurrent neural networks to model the visual activity. ii) They train the models in a full-supervised manner requiring manual annotations, which are tedious and expensive to acquire. iii) A severe limitation is that the visual activity in the facial gestures concerning other activities, such as laughing, chewing, etc., can be easily confused with the speech activity, thus adding false positives. 

To complement the noisy visual activity information, we propose to use the speaker identity information for active speaker face-voice assignment. Using the knowledge that face and voice both possess information about a speaker's identity, we present an unsupervised framework that exploits the fact that both the active speaker's face in the visual modality and the concurrent voice in the audio modality refer to the same speaker. We use well researched and readily available face and speech recognition systems to represent the faces and speech, respectively, so that both capture the speaker's identity. 

\begin{figure}
    \centerline{\includegraphics[width=0.45\textwidth,keepaspectratio]{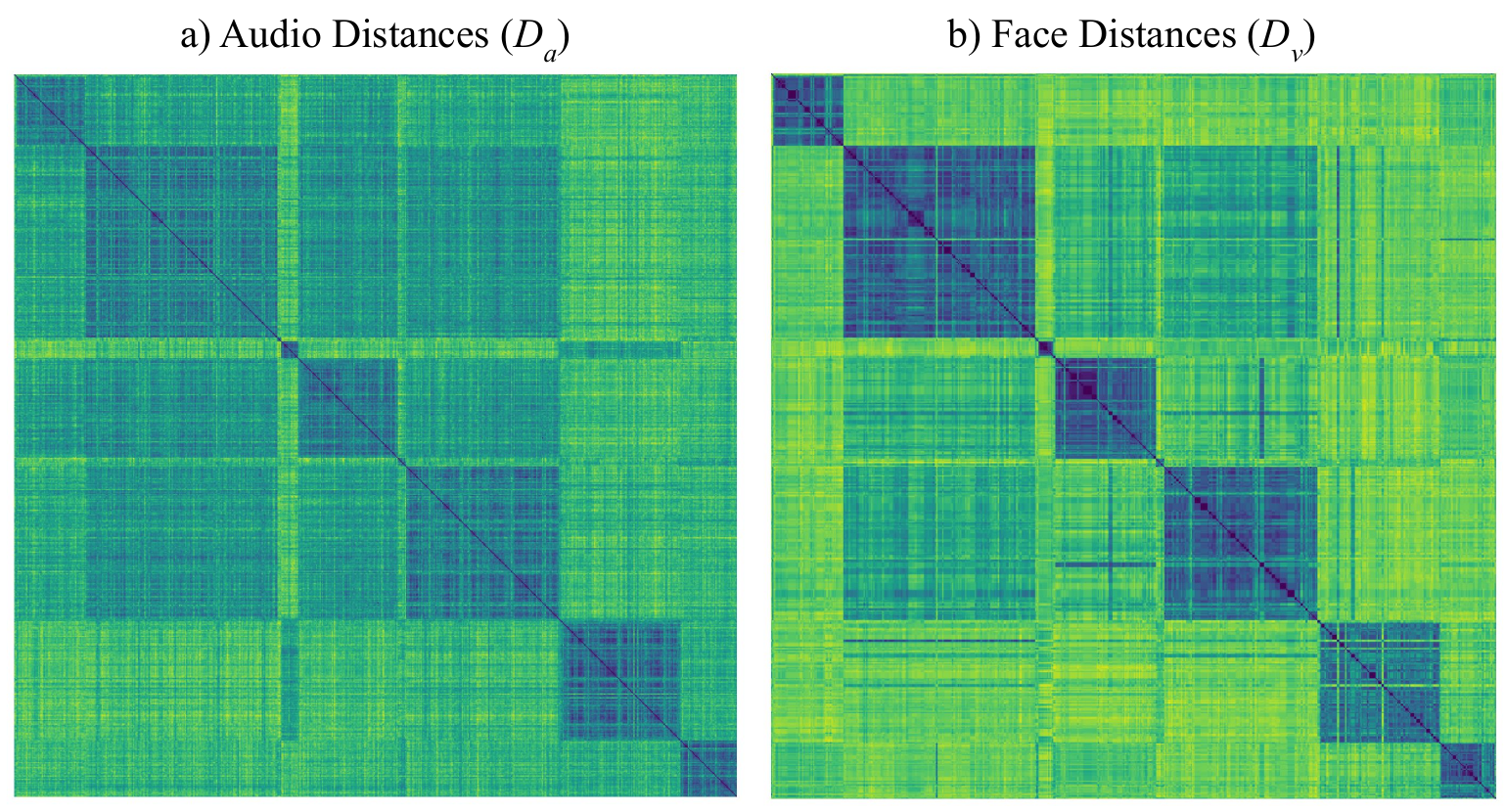}}
    \caption{a) Speech-identity distance matrix ($SD$) and b) Face-identity Distance matrix ($FD$) for the movie Hidden Figure. The active speaker faces are acquired form the ground truth. The speech segments are ordered to gather the speech segments of each speaker together.}
    \label{fig:gt_distances}
\end{figure}

\begin{figure*}
    \centerline{\includegraphics[width=\textwidth,keepaspectratio]{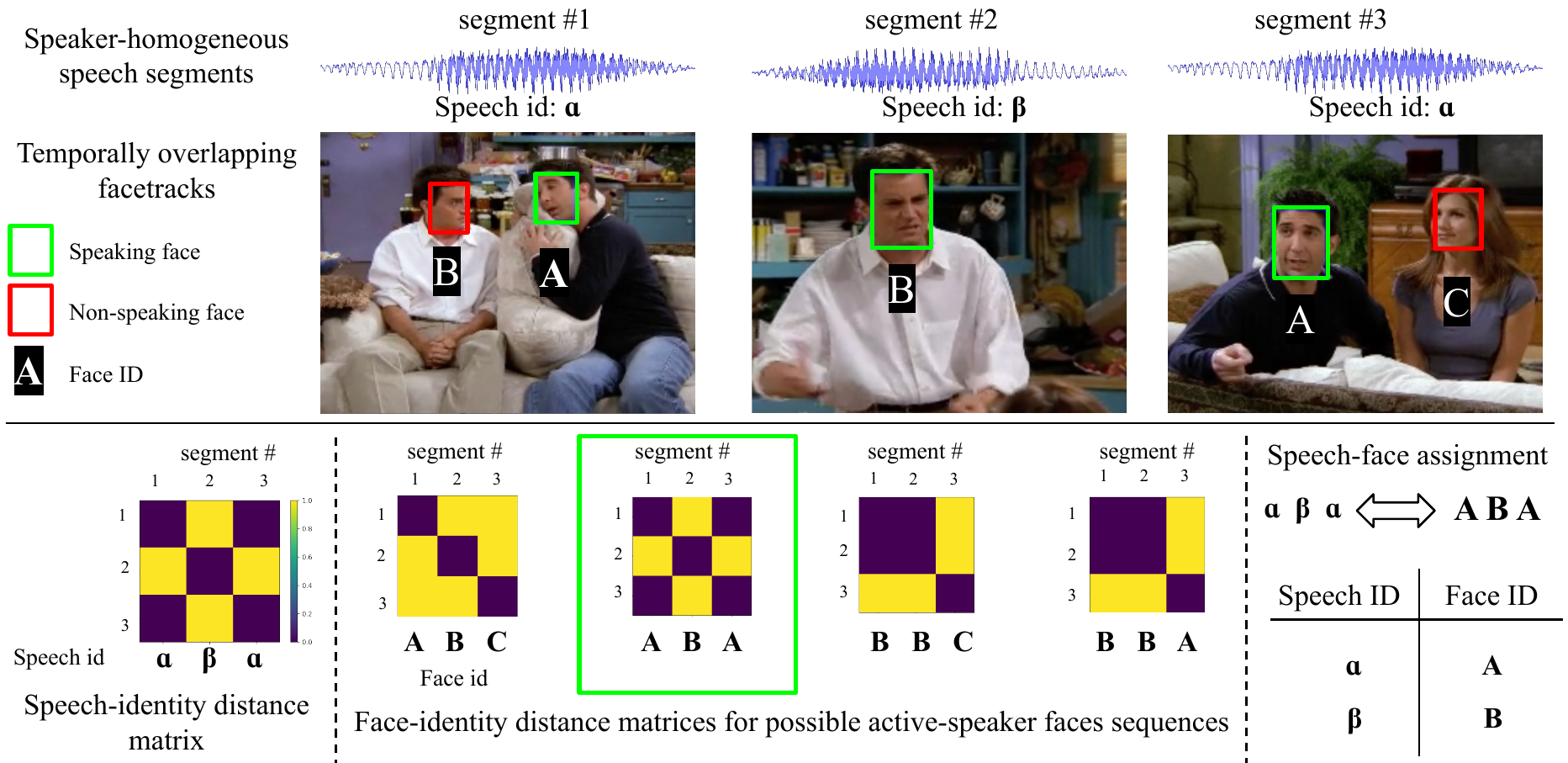}}
    \caption{The overview of the proposed framework: We gather temporally overlapping faces for each speaker-homogeneous speech segment. Using the speaker recognition embeddings, we construct a speech distance matrix. From the set of possible sequences of corresponding active speaker faces, we select the sequence of faces such that the face-identity distance matrix, obtained using the face-recognition features, displays a high resemblance with the speech-identity distance matrix.}
    \label{fig:overview}
\end{figure*}

Using the obtained representations, we construct a speech-identity distance matrix, capturing the relative identity structure of the speech segments through the entire video. Figure~\ref{fig:gt_distances}a shows an example of such a  distance matrix. We assign an active speaker face to each speech segment such that the obtained set of corresponding active speaker faces displays a similar relative identity structure in form of a face-identity distance matrix, an example is shown in Figure~\ref{fig:gt_distances}b. Figure~\ref{fig:overview} shows an overview of the proposed strategy. Furthermore, studying the similarity between the two distance matrices (speech and face), we present a simple strategy to address the speech segments with speakers present off-screen. 

We evaluate the system's performance for active speaker detection on three benchmark datasets: i) AVA active speaker~\cite{roth2020ava} and ii) Visual Person Clustering~\cite{brown2021face} datasets, consisting of videos of TV shows and movies, and iii) Columbia~\cite{chakravarty2016cross} dataset, consisting of a panel discussion video. The implementation code and the required dependencies to reproduce all the results in this work will be publicly available.  The contributions of this work are as follows:
\begin{enumerate}
    \item We present a novel framework to harness the speaker identity information, present within the speech in the audio modality and the speaker's face in the visual modality, for the speech-face assignment task. This framework can complement state-of-the-art visual-activity-based active speaker detection systems as the acquired information does not depend solely on the visual activity present in a frame.
    \item We formulate the active speaker detection task as a cross-modal unsupervised optimization task deriving information from pretrained speaker recognition and face recognition models. This eliminates the need for manual annotations and does not require additional training of any expensive models.
    \item We present an extensive performance evaluation of the system, validating the strategy's applicability for various challenging domains, including videos from TV shows, Hollywood movies, international movies, and panel discussions. Furthermore, we present ablation studies to study the performance of the proposed system and to highlight potential limitations. 
\end{enumerate}

\label{sec:methods}
    

%% file: related_work.tex
\section{Related Work}
\label{sec:related work}


An intuitive solution for detecting an active speaker is to study the activity in the lip region of faces in a video; this idea has led to a wide range of methods, from using just the visual signal~\cite{bendris2010lip} to posing it as multi-modal~\cite{owens2018audio,bendris2010lip, arandjelovic2018objects, chung2016out, chung2019naver} and cross-modal~\cite{chakravarty2015s,chakravarty2016cross,sharma2019icip,sharma2020cross} formulations. One of the earlier works proposed explicitly detecting the lip region and quantifying the observed activity to classify faces as talking or non-talking in TV shows~\cite{bendris2010lip}. Along similar lines,  sync-net~\cite{chung2016out} has proposed jointly modeling the activity in the explicitly detected lip region and the concurrent audio stream using a ConvNet architecture and training it to detect a speech-face synchronization leading to a self-supervised active speaker detection system. Following the success of modeling audio-visual synchronization, various self-supervised methods have emerged targeting the task of sound source localization~\cite{owens2018audio, arandjelovic2018objects, zhao2018sound, afouras2020self}. These methods were qualitatively shown to detect an active speaker's face as the location for the speech sound event.

The nature of speech activity, interleaving the audio and visual modalities, has led to various cross-modal approaches, predicting primarily audio-centric tasks such as voice activity detection~\cite{chakravarty2016cross,sharma2019icip, chakravarty2015s, sharma2020cross} using the visual input. Earlier works~\cite{chakravarty2015s, chakravarty2016cross} have modeled the facial activity using HoG features and trained SVM-based classifiers using weak supervision from audio VAD labels. More recently, the authors in ~\cite{sharma2019icip, sharma2020cross} introduced a more sophisticated 3D ConvNets architecture to model the visual frames and trained a weakly-supervised system for active speaker detection for the more challenging Hollywood movie video domain. Lately, the introduction of large-scale audio-visual datasets, such as AVA-active speaker~\cite{roth2020ava} and Active Speaker in Wild~\cite{kim2021look} consisting of manual active speaker annotations, has led to the emergence of a series of fully-supervised methods. These methods rely on the large-scale nature of labeled data to model the audio-visual activity by training train a wide variety of massive ConvNet architectures, ranging from 3D convolutions~\cite{chung2019naver, huang2020improved} to graph convolutional networks~\cite{alcazar2020active, leon2021maas}.

Unlike the extensively explored idea of modeling the visual activity,  efforts in the direction of utilizing the speaker identity information for active speaker detection is relatively limited. The work of Hoover et al.~\cite{hoover2018using} proposed to cluster the speech and faces separately to contain instances of a single speaker and then derive the information from the temporal co-occurrence of speech and face clusters to match speech and face for the same speaker. However, for the audio-visual diarization problem~\cite{park2022review}, the speaker's identity information from speech and faces is widely explored~\cite{kapsouras2015multimodal, bredin2016improving, chung2020spot}. For instance, using the temporal co-occurrence of the speech and the face identities (obtained using face clusters) to derive the speech identities~\cite{kapsouras2015multimodal}. Inspired by the same, in this work, we propose to use the well-established speaker recognition~\cite{indefense, voxceleb2} and face recognition~\cite{guo2016ms, senet} systems to capture the speaker's identity information from speech and faces and rely on speech and corresponding active speaker's face identifying the same character to derive speech-face associations.

%% file: methodology.tex
\section{Methodology}
\subsection{Problem Formulation}
The aim is to generate a speech-face association such that the associated face, if there is one, for every speech activity in the audio modality, is the source of the underlying speech. We formulate the problem definition under the following assumptions:
i) that the video content under consideration is shot to capture the speaker's face, which stands valid, especially in the case of entertainment media videos.
ii) that at any point in time, there can be at most one person speaking; thus, no overlapping speech. 
To investigate the validity of assuming no overlapping speech, we tabulate the fraction of overlapping-speech occurrences for several videos from VPCD and AVA active speaker datasets, in Table~\ref{tab:overlapping_speech}. We observe that the fraction is insignificant for all the videos, thus validating the assumption.

\begin{table}[]
\centering
\caption{Fraction of overlapping speech (\%) for various videos}
\label{tab:overlapping_speech}
\resizebox{0.3\textwidth}{!}{%
\begin{tabular}{@{}cc@{}}
\toprule
Videos & \begin{tabular}[c]{@{}c@{}}Fraction of \\ overlapping speech (\%) \end{tabular} \\ \midrule
Friends            & 1.39  \\
TBBT               & 4.42  \\
Sherlock           & 0.79  \\
Hidden Figures     & 0.44  \\
About Last Night   & 0.21 \\
AVA active speaker & 2.31  \\ \bottomrule
\end{tabular}%
}
\end{table}

We use the contiguous segments of speech activity in the audio modality of one speaker as the fundamental unit for our analysis. By definition, such a voiced segment consists of voices from only one speaker, called a speaker-homogeneous voiced segment. We denote the set of all the speaker-homogeneous segments through a video as \mbox{$S_{all} \equiv \{s_1 \dots  s_n, \dots s_N \}$}. Similarly, we use a face-track (locations of one face appearing in consecutive frames) as the smallest unit for the visual modality. For each $s_n$, we gather the temporally overlapping face-tracks available in visual modality and denote them as  \mbox{$F_n \equiv \{f_1, \dots f_k, \dots f_K\}$}. 

We call $f_k$ as the active speaker face for $s_n$, denoted as \mbox{$(s_n \longleftrightarrow f_k)$}, if face-track $f_k$ in video modality is the source of speech segment $s_n$ in audio modality. However, a speech segment, $s_n$, may have no face tracks available in the visual modality in its course, $F_n = \phi$; therein, the speaker is off-screen. Such segments are not relevant to this work, and hence purged from $S_{all}$. We only consider the speech segments which have one or more face-tracks overlapping in time, be the active-speaker's face is on-screen ($S_{\text{on}}$) or off-screen ($S_{\text{off}}$), denoted in equation \eqref{eq:background_def} and (2)
\begin{align}
\label{eq:background_def}
    S_{\text{on}} &\equiv \{s_n \mid  \exists f_k \in F_n : (s_n \longleftrightarrow f_k)\}\\
    S_{\text{off}} &\equiv \{s_n \mid F_n \neq \phi \text{ and } \nexists f_k \in F_n : (s_n \longleftrightarrow f_k)\}
\end{align}

We formulate the task of active speaker detection (ASD) as the problem of finding an active speaker face-track for all the speaker-homogeneous speech segments, if there exists one. We denote the set of active speaker faces, $A$, as:
\begin{align}
    A \equiv \{a_{n} \mid a_n = f_k \text{ and }  f_k \in F_n \text{ and } (s_n \longleftrightarrow f_k)\}
\end{align}
Such a formulation using the speaker-homogeneous speech segments helps reduce the ASD task to choosing one face track out of temporally overlapping tracks of (possibly multiple) faces for each speech segment. It also satisfies the constraint that active speaker faces can only be present when there is speech activity in the audio modality. At the same time, using a face track as one entity, rather than each face, helps facilitate that there can be at most one active speaker face in a frame which aligns with our assumption of non-overlapping speakers. This formulation intends to find an active speaker face for each speech segment, $s_n$, but as mentioned earlier, the speech segments may have off-screen speakers ($s_n \in S_\text{off}$), which may introduce false positives. 

Since the information separating the speech segments into the sets $S_{\text{on}}$ and $S_{\text{off}}$ is not readily available we propose a two-stage process. We first devise a system that provides an associated active speaker face-track for all speech segments in $S_a \equiv S_{\text{on}} \cup S_{\text{off}}$ and call it \emph{stage1}. This is then followed by a system to remove the speech segments with off-screen speakers from the set $S_\text{all}$, to correct the introduced  errors and call it \emph{stage2}. 

\subsection{Stage1: Speech-face Assignment}
\label{methods:stage1}
Speech in the audio and face in the visual modality is widely known to possess an individual character's identity information. Utilizing it, we propose to select an active speaker face track for each speech segment $s_n$ from the set of temporally overlapping face tracks, $a_n \in F_n$, imposing that the selected face and the underlying speech depict the same character identity. Since the identity captured in the audio modality (using speech) can not be directly compared with the captured identity in the visual modality (using faces), we compute an identity distance matrix for the two modalities independently, bringing them into a shared space of relative identities. 

We start with initializing the set of active speaker face tracks $A$ by randomly selecting a face track for each speech segment from the set of face tracks temporally overlapping with the speech segment,  $a_n = f_k \mid f_k \in F_n \forall s_n \in S_{all}$. Then we construct representations for the speech segments and the active speaker faces to capture the character identity information. For speech segments, we extract the speaker recognition embeddings using a ResNet34~\cite{resnet} pretrained on VoxCeleb2~\cite{voxceleb2} with angular learning objective~\cite{indefense}. We represent the faces $f_{k}$ using SENet-50~\cite{senet}, pretrained on MS-Celeb-1M~\cite{guo2016ms}, widely used in face recognition systems.

Using the obtained representations, we construct a distance matrix for each modality. In the audio modality, we represent the speech segments, $s_n$ in terms of its distances from all other speech segments. We compute the distance between two speech segments $s_i$ and $s_j$ as the cosine distance between their speaker recognition embeddings. We call this matrix as the speech-identity distance matrix, $SD$ and show it in equation~\ref{eq:disatance_matrix}. Similarly, in the visual modality, we construct a face-identity distance matrix, $FD$ representing each active speaker face, $a_n$ in terms of its distance from all other active speaker faces. We compute the distances between two active speaker faces $a_i$ and $a_j$ as the cosine distance between their obtained face recognition embeddings, as shown in equation~\ref{eq:disatance_matrix}. While computing these distance matrices, we preserve the order of the speech segments and corresponding active speaker faces, implying that the $i^{th}$ row of $SD$ and $FD$ represents the speech segment, $s_i$, and the corresponding active speaker face, $a_i$, respectively.

\begin{align}
\label{eq:disatance_matrix}
    SD[i,j] &= \frac{s_{i} \cdot s_{j}}{\|s_{i}\| \|s_{j}\|} &
    FD[i,j] &= \frac{a_{i} \cdot a_{j}}{\|a_{i}\|\|a_{j}\|}
\end{align}

The use of identity-capturing speaker recognition representations for speech segment, $s_n$ enables the distance matrix $SD$ to capture the relative identity structure with context from the entire video. For instance, speech segments $s_i$ and $s_j$ from the same speaker will show a smaller value in $SD[i,j]$ than those from different speakers. Similarly, the face-identity distance matrix $FD$ constructed using the face recognition representations captures the relative identity of the active speaker faces in the visual modality. Figure~\ref{fig:gt_distances} shows the $SD$ and $FD$ for the Hollywood movie \emph{Hidden Figures}, where we have constructed the matrix $FD$ using the ground truth active speaker faces. For the visualization purposes, we gather the speech segments of each speaker together. Since the active speaker's face and the concurrent speech must identify the same person, we hypothesize that the relative identity structure captured by $FD$ should have a high resemblance with $SD$. We observe this similarity in Figure~\ref{fig:gt_distances}, in the form of similar low-distance square formations along the diagonals of both the matrices.

The $i^{th}$ row in both the matrices depicted as $SD[i]$ and $FD[i]$ represents the identity of the underlying person (character) in terms of its distance from all other characters in the video. Since the components of the two rows come from distances computed in two independent dimensions (the audio and visual modality), they may differ in scale and thus may not be directly comparable. However, as the components of each row vector show the distance relative to other identities and all the identities are supposed to be the same among the two matrices, we expect the row vectors from the two matrices to show a linear relationship. So we propose to use Pearson's correlation as the measure to quantify the similarity between the two matrices. We formalize the ASD task as an optimization problem where we select the active speaker faces for each speech segment from the set of temporally overlapping face tracks, such that the Pearson's correlation between the speech-identity matrix and face-identity matrix, computed as an average of the row-wise correlations, is maximized. The formulation is denoted below in equation~\ref{eq:optimization}, where $SD[i]$ and $FD[i]$ denotes the vector corresponding to the $i^{th}$ rows of $SD$ and $FD$ respectively and $N$ is the number of rows.

\begin{equation}
\label{eq:optimization}
  \begin{split}
      Corr(FD) =\\
      \frac{1}{N}\sum_i&\frac{\sum (SD[i] - \overline{SD[i]})(FD[i] - \overline{FD[i]})}{\sqrt{\sum(SD[i] - \overline{SD[i]})^2\sum(FD[i] - \overline{FD[i]})^2}}
  \end{split}
\end{equation}
\begin{equation}
     A \equiv \argmax_{\{ a_n \mid a_n = f_{k} \text{ and } f_k \in F_n\}}Corr(FD)
\end{equation}

The notable point is that the face-identity distance matrix, $FD$ is symmetric; the rows are not independent. The change in the active speaker's face for $i^{th}$ speech segment implies the change in the $i^{th}$ row of $FD$, which reflects a change in all the other rows of $FD$ in the form of the altered $i^{th}$ column. Changed $FD$ further reflects a change in the objective function $Corr(FD)$. Thus, in the ideal case scenario, the active speaker faces for all the speech segments should be jointly selected to maximize the $Corr(FD)$. Without loss of generality, let us consider that there are $N$ speech segments in a video and, on average, $k$ faces that temporally overlap each speech segment. The complexity for computing the objective function $Corr(FD)$ for a given set of active speaker faces, $A$ would be $O(N^2)$, and there would exist $k^N$ such possible sets, $A$ (all possible combinations of potential active speaker faces). Thus maximizing the $Corr(FD)$ jointly for all active speaker faces will cost $O(k^NN^2)$. For a typical TV Hollywood movie, $k \approx 3$ and $N \approx900$ bring the cost of jointly optimizing high and out of the scope of current computing standards. 

We propose to approximate the optimization process by iteratively finding the active speaker's face corresponding to one speech segment at one time to maximize the objective function $Corr(FD)$, keeping the active speaker faces for all other speech segments fixed. Once iterating through all the speech segments (an epoch), we repeat the process until the $Corr(FD)$ converges. This approximation reduces the complexity of the optimization from an exponential to a cubic order of $N$, $O(EkN^3)$, where E is the number of epochs, $kN$ is the number of times the $Corr(FD)$ is computed in each epoch and $O(N^2)$ is complexity for computing $Corr(FD)$. Furthermore, assuming that even a portion of a movie has enough information to disambiguate between the speech and speaker's faces, we propose to split the number of speech segments into $p$ partitions and optimize them separately. This helps further reduce the complexity of the process to $O(Ek\frac{N^3}{p^2})$ where $p>1$. The process is detailed in Algorithm~\ref{algo:optimzation}.  

\begin{algorithm}
\label{algo:optimzation}
$\text{Random initialization:} A \equiv \{a_n \mid  a_{n} \in F_{n}\} \forall s_n \in S_\text{all} $\;
Compute $SD$ and $FD$ \tcp*{using eq:~\eqref{eq:disatance_matrix}}
$objective \gets Corr(FD)$ \tcp*{using eq:~\eqref{eq:optimization}} 
\While{$objective$ increases}{
    \For{each $a_i \in A$}{
    $a_{i} = \argmax\limits_{f_k \in F_{n}} Corr(FD)$ \tcp*{$i^{th}$ row}
    $\text{Update } FD$\;
    $\text{Update } A$\;
    }
    $ \text{$objective$} \gets Corr(FD)$\;
  }
  \caption{Stage1: Speech-face assignation}
\end{algorithm}
\subsection{Stage2: Off-screen speaker correction}
\label{sec:background_method}

Maximizing the objective function, $Corr(FD)$, we assign a corresponding active speaker face for all the speech segment under consideration, i.e., $s_n \in S_\text{all}$, which includes an active speaker face even for the speech segments having off-screen speakers, $s_n \in S_\text{Off}$. This incorrect assignment of speakers' faces introduces false positives.

The underlying hypothesis to maximize the correlation between the speech-identity and the face-identity distance matrices is that since the active speaker's faces and the speech segments must identify the same person, the relative identity patterns from the speech-distance matrix ($SD$) and the face-distance matrix ($FD$) should show a high resemblance. In contrast, in the case of speech segments with speaker present off-screen, none of the face tracks temporally overlapping with the speech segment corresponds to the active speaker's face. Thus the speech segment represents an identity different from any of the concurrent face tracks. We hypothesize that this discrepancy in the identity leads to a relatively lower similarity between the identity representations obtained using the underlying speech segment and any of the temporally overlapping faces for the speech segments with off-screen speakers. Formally we classify the speech segment, $s_i$, as the speech segment with an off-screen speaker if it displays a low enough row-wise correlation between the identity representations obtained from the speech-distance matrix, $SD[i]$ and any of the temporally overlapping face tracks, $FD[i]$, denoted in Eq~\ref{eq:offscreen}. Furthermore, we remove the face tracks, earlier incorrectly classified as active speaker faces corresponding to such speech segments with off-screen speakers. 

\begin{equation}
\label{eq:offscreen}
S_\text{off} \equiv \left\{ s_i \big | \frac{\sum (SD[i] - \overline{SD[i]})(FD[i] - \overline{FD[i]})}{\sqrt{\sum(SD[i] - \overline{SD[i]})^2\sum(FD[i] - \overline{FD[i]})^2}} < \tau \right\}    
\end{equation}

%% file: experiments.tex
\section{Performance Evaluation}
\label{sec:experimetns}
\subsection{Implementation details}

We first obtain the active voice regions in a  video using an off-the-shelf speech segmentation tool called pyannote~\cite{pyannote}. As the proposed system requires speaker homogeneous speech segments, we use naive heuristics to segment further the obtained voice active regions instead of employing a sophisticated speaker change detection system. We partition the voiced regions at the shot boundaries, which we gather using PyScenedetect\footnote{http://scenedetect.com/en/latest/}. The partitioning at the shot boundary is motivated by speaker change being a prominent movie cut attribute~\cite{pardo2021moviecuts}, thus decreasing the likelihood of observing a speaker change in the obtained segments. We further partition the obtained segments to have a maximum duration of 1sec, effectively decreasing the fraction of speech segments consisting of a speaker change. We represent the speech segments using a ResNet-34~\cite{indefense}  model pretrained on Voxceleb2~\cite{voxceleb2} dataset for the speaker recognition task.

Within the gathered shot boundaries, we use the RetinaFace~\cite{deng2020retinaface} to obtain face detections and track them using the SORT tracker~\cite{Bewley2016_sort} to construct face tracks. We extract identity representations for each face detection box using a SENet-50~\cite{senet}, pretrained on MS-Celeb-1M~\cite{guo2016ms} for the face recognition task, and average them over all the constituting faces to construct a representation for a face track. Once the representations for all the speaker homogeneous speech segments and all the faces are gathered, we initialize the active speaker face for each speech segment with a face randomly selected from a set of temporally overlapping faces. We partition the set of speech segments to contain at most 500 speech segments (L=500), grouped in temporal order, which helps reduce the optimization process's time complexity. We then employ the Algorithm~\ref{algo:optimzation} to assign an appropriate active speaker face to each of the speech segments for each partition (stage1) and further improve the predictions by discovering the speech segments with off-screen speakers (stage2).

To evaluate the performance of the proposed system, we use three benchmark datasets: i) Visual person clustering dataset (VPCD)~\cite{brown2021face}, ii) AVA-active speaker dataset~\cite{roth2020ava} and iii) Columbia dataset~\cite{chakravarty2016cross}. The AVA-active speaker dataset consists of face-wise active speaker annotations for 161 international movies, freely available on YouTube. We use the publicly available validation split, consisting of 33 movies, to report and compare the performance of the proposed strategy against the state-of-the-art systems. VPCD is more exhaustive, consisting of character identity information in  addition  to the active speaker annotations for videos from several widely watched episodes of TV shows (Friends, Sherlock and TBBT) and 2 Hollywood feature films (Hidden Figures and About Last Night). It accompanies bounding boxes for the appearing faces and their corresponding character identities, along with the information about which character is speaking at any time. One of the challenges of working with VPCD is that the video files are not freely available. For purposes of this research, we acquired the video files using the DVDs and aligned the annotations in VPCD with the acquired videos. The Columbia dataset consists of manually obtained active speaker annotations for an 85 minutes long video of a panel discussion with six speakers. The panel discussion scenario is relatively more controlled than the movies in terms of frontal face poses and noise-free speech.

\begin{table}[]
\centering
\caption{F1-scores at various stages for the videos from VPCD. Reported F1 scores are averaged over all the episodes for the TV shows (Friends, TBBT, and  Sherlock).}
\label{tab:f1-scores}
\resizebox{0.5\textwidth}{!}{%
\begin{tabular}{c|c|c|c|c}
\toprule
Videos                & \begin{tabular}[c]{@{}c@{}}Random \\ initialization\end{tabular} & \begin{tabular}[c]{@{}c@{}}Stage 1\\ speech-face \end{tabular} & \begin{tabular}[c]{@{}c@{}}Stage 2\\ Off-screen speakers\end{tabular}  & \begin{tabular}[c]{@{}c@{}}Stage 2\\(mAP)\end{tabular}\\ \midrule
Friends (25 episodes) & 0.54                                                             & 0.78                                                              & 0.80         &0.82                                                         \\ 
TBBT (6 episodes)     & 0.60                                                             & 0.83                                                              & 0.84               &0.86                                                   \\ 
Sherlock (3 episodes) &0.55                                                                  &0.64                                                                   &0.66         &0.70                                                                    \\ 
Hidden Figures      & 0.45                                                             & 0.62                                                              & 0.67     &0.66                                                             \\ 
About Last Night        & 0.51                                                             & 0.59                                                              & 0.62   &0.57                                                               \\ \bottomrule
\end{tabular}%
}
\end{table}
\subsection{Stage1: Speech-face assignation}
\begin{figure*}[]
    \centering
    \includegraphics[width=\textwidth,keepaspectratio]{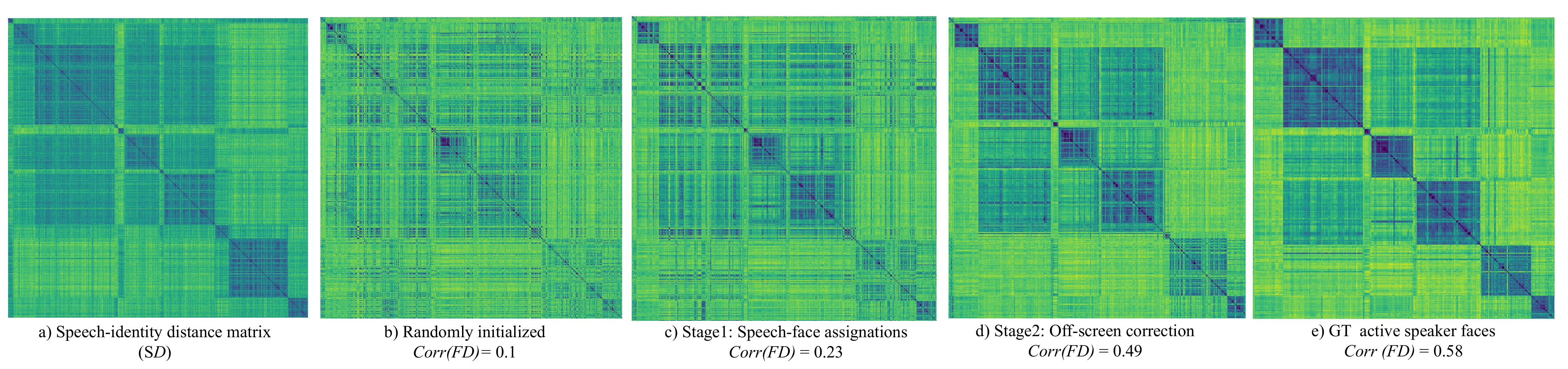}
    \caption{Distance matrices for the movie Hidden Figures at various stages of the system along with the value of objective function, $Corr(FD)$. a) Speech-identity distance matrix ($SD$) b)Random: Face-identity distance matrix ($FD$) for randomly initialized ASD. c) Stage1: $FD$ with speech-face assignation, maximizing $Corr(FD)$. d) Stage2: $FD$ post removing the speech segments with off-screen speaker. e) Ground truth: $FD$ obtained using ground truth active speaker faces.}
    \label{fig:distance_matrices}
\end{figure*}

The hypothesis underlying the Algorithm~\ref{algo:optimzation} is that the relative identity patterns captured by the face-identity distance matrix $FD$, constructed using the active speaker faces, must show a high resemblance with the speech-identity distance matrix $SD$. In this section, we present the qualitative and quantitative validation of the hypothesis using videos from the VPCD. We compare the predictions for each face-bounding box against the ground truth and report the F1-score in Table~\ref{tab:f1-scores}. 
In Figure~\ref{fig:distance_matrices} we show the evolution of the $FD$ at various stages for the movie Hidden Figures. 

Using ground truth speaker identities, we order the speech segments in $SD$, and the corresponding predicted active speaker faces in $FD$ to bring the speech segments from each speaker together. This makes the matrices interpretable and brings out the visible square patterns along the diagonal in $SD$ in Figure~\ref{fig:distance_matrices}a. Figure~\ref{fig:distance_matrices}b shows the $FD$ for the random initialization case with a low value of the $Corr(FD)$, displayed in the form of minutely visible similarity with $SD$. However small, the visible similarity signifies the correct predictions for the trivial case of speech segments with just the speaker's face visible in the frames. The performance of the random chance system is quantified in terms of F1-scores in Table~\ref{tab:f1-scores}. 

\begin{figure}[]
    \centering
    \includegraphics[width=0.45\textwidth,keepaspectratio]{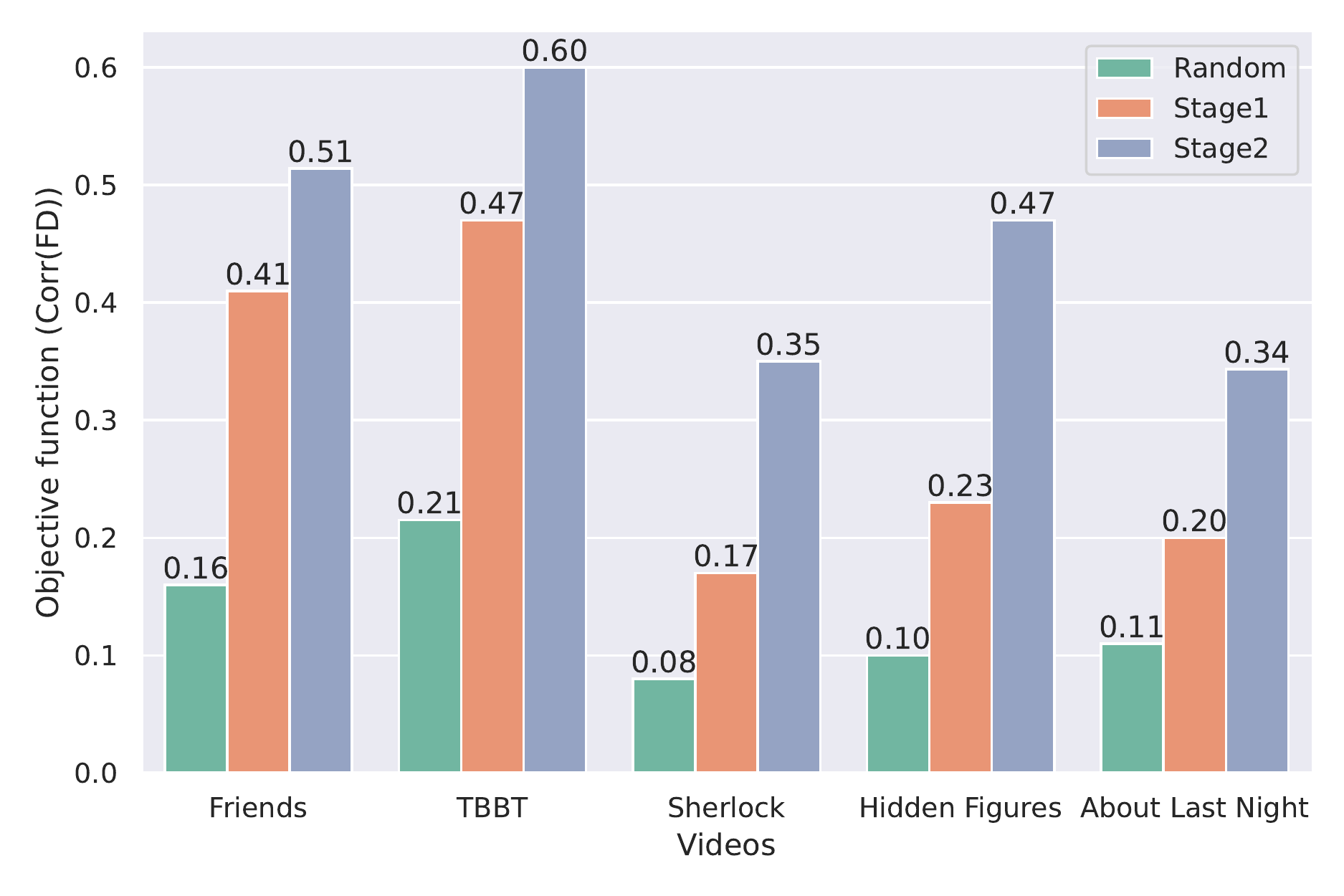}
    \caption{Evolution of the objective function $Corr(FD)$ at different stages for the videos in VPCD.}
    \label{fig:correlations}
\end{figure}

Figure~\ref{fig:distance_matrices}c shows the $FD$ post stage1 speech-face assignments, maximizing the $Corr(FD)$ employing Algorithm~\ref{algo:optimzation} and displays significantly higher resemblance with $SD$, quantified by the higher value of $Corr(FD)$. Figure~\ref{fig:correlations} compares the objective function value ($Corr(FD)$) at various stages for the videos in VPCD, and shows the higher value fro stage1 than the random initialization case, consistently for all the videos. The higher resemblance translates to the enhanced active speaker detection performance, reported in terms of a higher F1-score in Table~\ref{tab:f1-scores}, compared to the random chance system. We consistently observe a substantially higher F1-score than the random chance baseline for all the videos in the VPCD. Although post optimization, the $FD$ is visibly similar to $SD$ (comparing Figure~\ref{fig:distance_matrices}a\&c), the amount of noise is still notable when compared against the $FD$ obtained using the ground truth active speaker faces, shown in Figure~\ref{fig:distance_matrices}e. The same is quantitatively visible in the difference between respective values of $Corr(FD)$ and is due to the false positives introduced by the system for the speech segments with off-screen speakers.  

\subsection{Stage2: Off-screen speaker correction}
The proposed strategy to segregate the speech segments with off-screen speakers relies on the hypothesis that for such speech segments, since the speaker's identity from the speech is different from any of the faces appearing in the frames, the row-wise correlations between $SD$ and $FD$ are relatively lower. To validate this hypothesis, we first divide the set of speech segments, using the ground truth, into $S_\text{On}$ (on-screen) and $S_\text{Off}$ (off-screen) and study the distribution of row-wise correlations between $SD$ and $FD$ for the two groups. We assign ground truth active speaker face for $s_i \in S_\text{On}$ and a face randomly selected from the set of temporally overlapping face tracks for $s_i \in S_\text{Off}$ and then construct the $SD$ and $FD$ using all speech segments $s_i \in S_\text{On} \cup S_\text{Off}$. 

\begin{figure}[tb]
    \centering
    \includegraphics[width=0.5\textwidth,keepaspectratio]{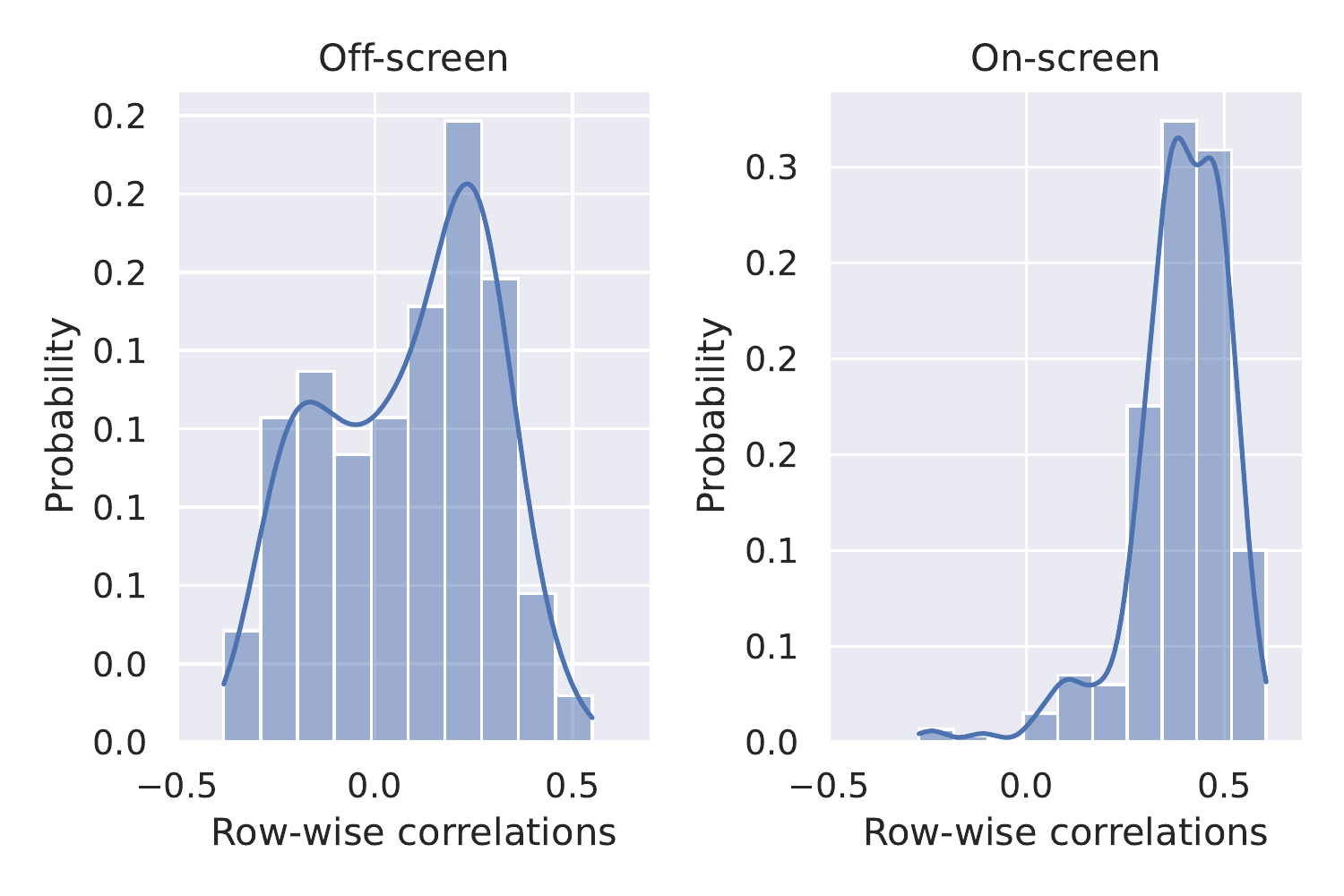}
    \caption{Plot showing the individual contribution ($corr_i$) to $O(D_v)$ vs the change in contribution ($\Delta_i$) when substituted with other character faces, for speech segments a) having background speakers ($S_b$), b) having on-screen speakers ($S_o$).}
    \label{fig:background_corr}
\end{figure}

In Figure~\ref{fig:background_corr}, we show the distributions of row-wise correlations between $SD$ and $FD$ for the two sets of speech segments: i) Off-screen, $S_\text{Off}$ and ii) On-screen, $S_\text{On}$, for the movie Hidden Figures. We observe a notable difference in the two distributions, which we quantify using the non-parametric Mann-Whitney U~\cite{mann1947test} test and obtain a p-value of $10^{-70}$ verifying that the difference between the two distributions is statistically significant. We performed the same test for all other videos in VPCD and found that the hypothesis held for all of them. 

In Table~\ref{tab:auROC} we report the fractional duration of speech segments having off-screen speakers ($S_\text{Off}: S_\text{all}$) for each video in the VPCD and observe that the fraction can be significant for some videos and thus validating the need to address the induced errors in active speaker detection by the stage1 speech-face assignations. Using various values of $\tau$ in equation~\ref{eq:offscreen}, we compute an ROC curve to evaluate the efficacy of the proposed strategy to classify each speech segment into $S_\text{Off}$ and $S_\text{On}$. We report the area under ROC (auROC) curve in Table~\ref{tab:auROC} for all the videos and notice the good performance of the simple strategy to discover speech segments with off-screen speakers. 

\begin{table}[]
\centering
\caption{Performance for the off-screen speech segment classification for the videos in VPCD, in terms of area under the ROC curve (auROC).}
\label{tab:auROC}
\resizebox{0.45\textwidth}{!}{%
\begin{tabular}{ccc}
\toprule
Videos           & \begin{tabular}[c]{@{}c@{}}off-screen \\ speech segments (\%)\end{tabular} & \begin{tabular}[c]{@{}c@{}}Classification\\ performance (auROC)\end{tabular} \\ \midrule
Friends          & 0.1                                                                        & 0.8074                                                                       \\ 
TBBT             & 0.1                                                                        & 0.8057                                                                       \\ 
Sherlock         & 0.18                                                                       & 0.6927                                                                       \\ 
Hidden Figures & 0.21                                                                       & 0.8259                                                                       \\ 
About Last Night   & 0.17                                                                       & 0.6897                                                                       \\ \bottomrule
\end{tabular}%
}
\end{table}

For all the videos in VPCD, we use $\tau=0.1$ to gather the speech segments with off-screen speakers and classify the faces, earlier incorrectly marked as active speaker faces in stage1, as non-speaking faces. In Figure~\ref{fig:distance_matrices}d, we show the obtained $FD$ for the movie, Hidden Figures and observe a visibly higher similarity with the speech-identity distance matrix ($SD$), further quantified in attained a higher value of $Corr(FD)$, compared to $FD$ post stage1.   We further observe that the  $FD$ post stage2 looks even more similar to ground truth $FD$. 
In Figure~\ref{fig:correlations}, we show that the objective function $Corr(FD)$ attains a higher value, after correcting the speech-face assignments for speech segments with off-screen speakers, compared to stage1 assignments for all the videos in VPCD.  

\begin{figure}[]
    \centering
    \includegraphics[width=0.45\textwidth,keepaspectratio]{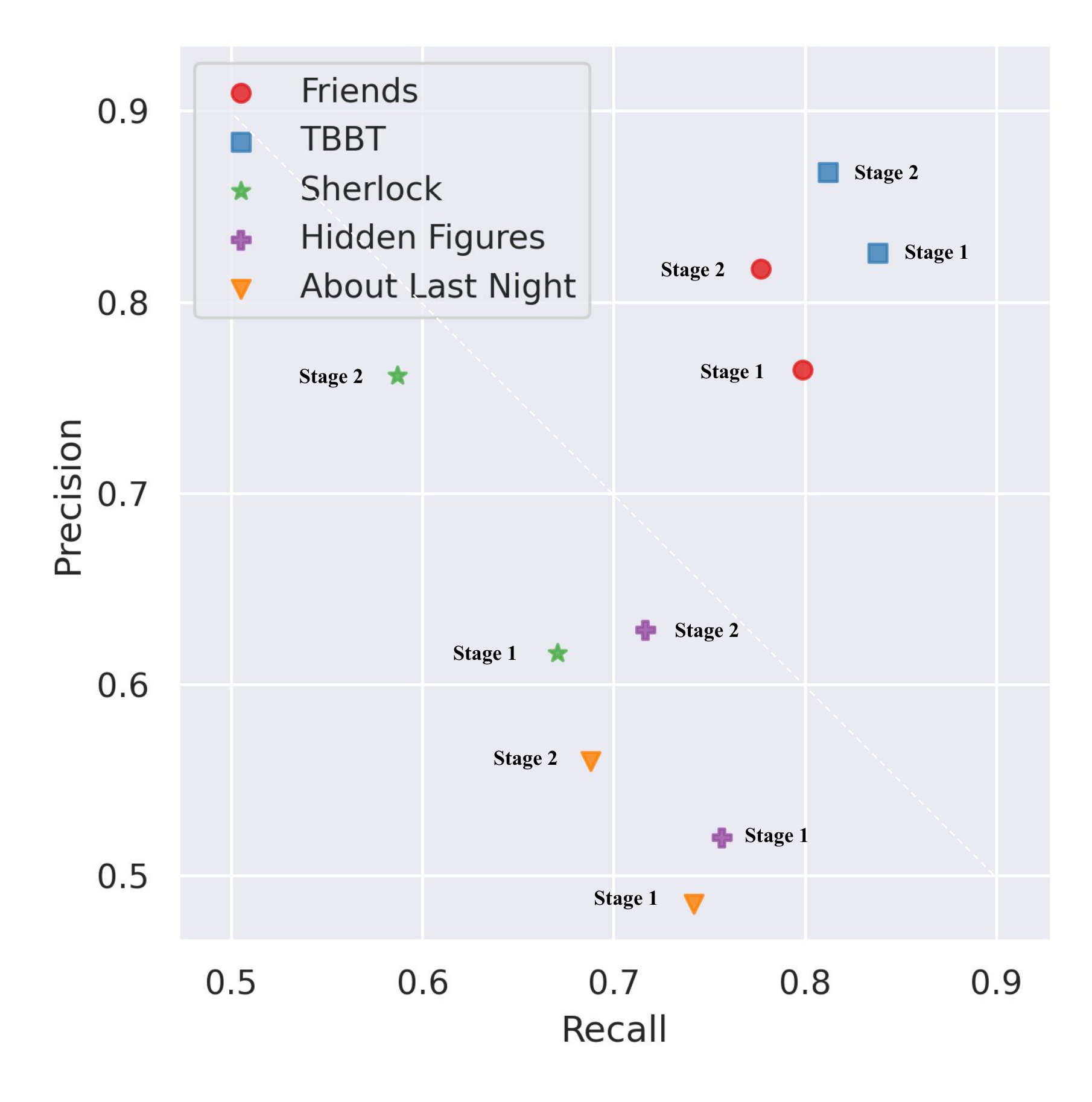}
    \caption{Increase in precision with a slight drop in recall introduced by correcting the speech-face assignments for speech segments with off-screen speakers (stage2). }
    \label{fig:stage1v2}
\end{figure}

We report the F1-score for the active speaker detection using the corrected predictions in Table~\ref{tab:f1-scores} (stage2) and compare them with the predictions from the stage1. We observe that off-screen speaker correction increases the F1-score consistently for all the videos. In Figure~\ref{fig:stage1v2}, we compare the performance of the two stages on a precision-recall curve and observe that the system's precision improves significantly with a slight drop in the recall. The enhanced precision is associated with removing the false positives introduced by assigning an active speaker face to speech segments with off-screen speakers (stage1), while the decline in the recall is due to the incorrect classification of a speech segment to have an off-screen speaker in stage2.

\subsection{Comparisons with state-of-the-art}
We evaluate the performance of the proposed strategy involving finding an active speaker face track for each speech segment and then removing the speech segments with off-screen speakers on the videos from the publicly provided validation split of the AVA dataset. We score the face boxes of the predicted active speaker face tracks with the row-wise correlation ($correlation(SD[i], FD[i])$) values where a higher correlation value signifies a better match in identities represented by the face and the speech. All other face boxes are scored -1, the minimum possible correlation value. We then use the AVA official implementation to evaluate the performance and report the mean average precision (mAP) values aggregated over all the videos in Table~\ref{tab:AVA}. We also report the performance of various recent state-of-the-art techniques in Table~\ref{tab:AVA}. 

\begin{table}[]
\centering
\caption{Comparison with the state-of-the-art methods on AVA active speaker validation set in terms of mean average precision.}
\label{tab:AVA}
\resizebox{0.35\textwidth}{!}{%
\begin{tabular}{@{}ccc@{}}
\toprule
Methods           & Stragetgy             & mAP (\%)       \\ \midrule
Roth et al.~\cite{roth2020ava}       & Supervised            & 79.2           \\
Zhang et al.~\cite{zhang2019multi}       & Supervised            & 84.0           \\
Alxazar et al.~\cite{alcazar2020active}    & Supervised            & 87.1           \\
Chung et al~\cite{chung2019naver}       & Supervised            & 87.8           \\
MAAS-TAN~\cite{leon2021maas}        & Supervised            & 88.8           \\ \midrule
Proposed          & Unsupervised          & 71.40 \\ \bottomrule
\end{tabular}%
}
\end{table}

We observe that the proposed unsupervised strategy of using the identity information performs well, although not as well as other fully supervised methods using videos from the AVA train set to model the visual activity in the frames. The primary reason for the observed lower performance lies with the nature of videos in the AVA dataset which consists of adverse audio conditions. Nearly 65\% of the speech segments with a visible speaker overlap with either noise or music~\cite{roth2020ava}. The presence of such noise conditions hinders the ability of the employed speaker recognition models to capture the identity information from speech segments, which naturally translates into poor active speaker detection performance. We further elaborate on the effect of speaker recognition performance on active speaker detection in section~\ref{ablation:speech_recognition}.

Using the same method as AVA, we report mAP values for the videos in the VPCD dataset (averaged over episodes for TV shows) in Table~\ref{tab:f1-scores}.
We observe that the performance for videos of TV shows, specifically Friends and TBBT, is notably higher than the movies (Hidden Figures and About Last Night). The scenes in Friends and TBBT predominantly consist of just primary characters' faces (characters who actively speak in the video), unlike the movies, which have a significant portion of scenes consisting of faces of background characters (i.e., characters who do not speak anytime in the video). The presence of primary characters' faces provides useful disambiguating information since we can gather their audio-visual identity information from the parts of videos where they speak. While the background characters'  faces contribute to increased confusion since there is no way we can acquire their identity information. This leads to the displayed difference in the performances for the TV shows and the movies. Additionally, a higher fraction of off-screen speakers for the movies than the TV shows further contributes to the lower performance for movies (Table~\ref{tab:auROC}). Selected video clips, presenting the system's active speaker predictions for several videos from VPCD are present in supplementary material. 

\subsection{Ablation studies}
\subsubsection{\textbf{Effects of VAD performance}}
The availability of the speaker-homogeneous speech segments is fundamental to the proposed approach. To obtain the speech segments, we use an off-the-shelf voice activity detector~\cite{pyannote}. To make the obtained speech segments speaker-homogeneous, we use a simple approximation involving partitioning the speech segments by the scene boundaries and ensuring a maximum duration of 1sec. Since the number of speaker changes is constant in a video, splitting the speech segments into a larger number of partitions (obtained by making the partition length smaller) effectively decreases the fraction of speech segments consisting of a speaker change. Furthermore, the errors introduced by the speech segments with a speaker change can be the most for 1sec duration. 

\begin{figure}[]
    \centering
    \includegraphics[width=0.45\textwidth,keepaspectratio]{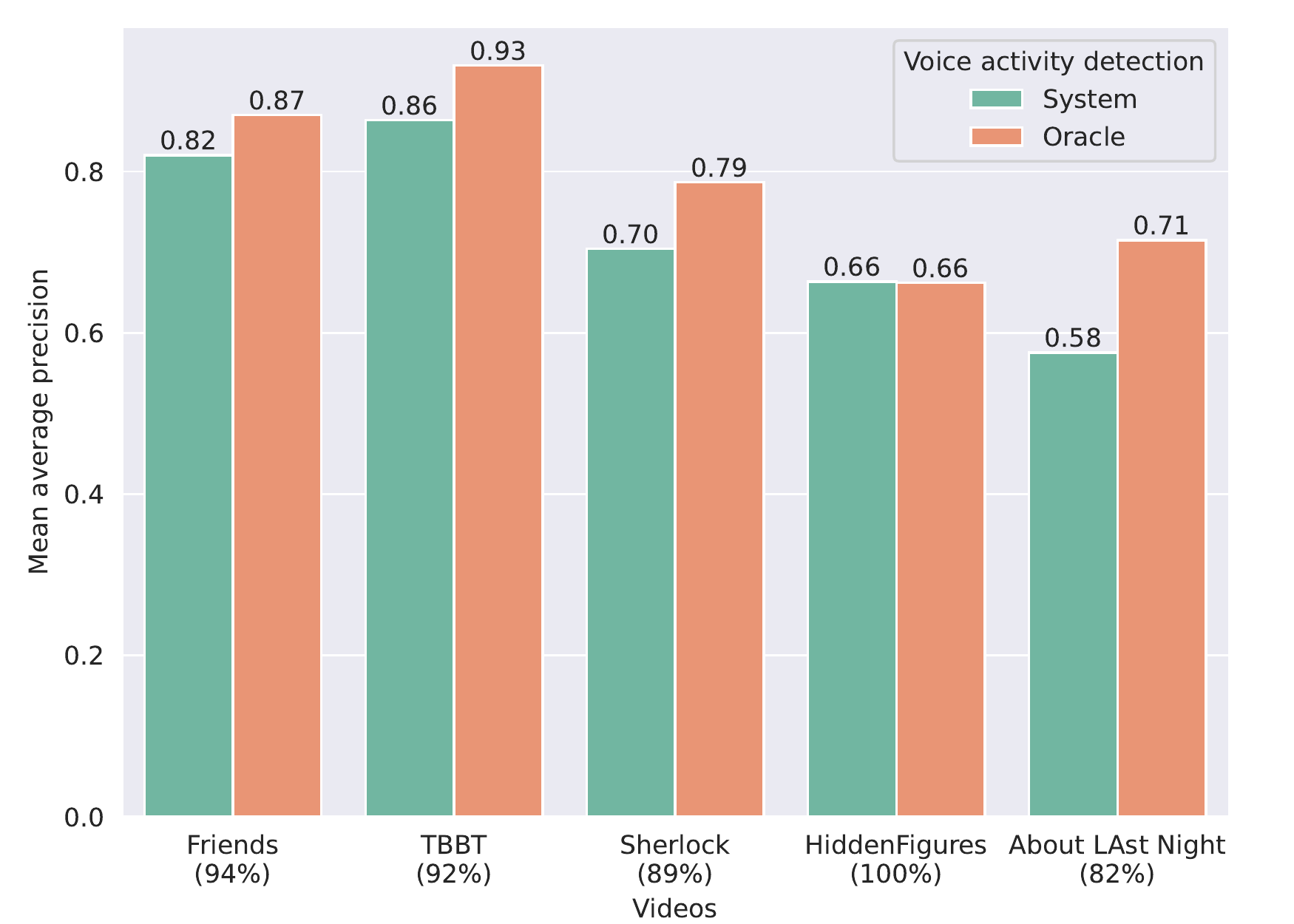}
    \caption{Comparison of the system's performance against the scenario with ideal case speaker-homogeneous speech segments, reported in terms of mAP for video from VPCD. The system's performance relative to the ideal case scenario is shown beneath each video.  }
    \label{fig:gtVAD}
\end{figure}

To evaluate the effect of the proposed simple proxy for speaker-homogeneous speech segments, we compare the system's performance, on videos from VPCD, with the ideal case scenario. The VPCD annotations consist of character-wise speech segments, enabling to obtain ideal case speaker-homogeneous speech segments. We further segment the speech segments using the scene boundaries and ensuring a maximum duration of 1 sec. As the speech recognition methods' performance is susceptible to the duration of segments, splitting to maximum duration of 1 sec eliminates the impact of the difference in durations of speech segments in the two cases. In Figure~\ref{fig:gtVAD} we compare the performance in terms of mAP for all the videos in VPCD for the two cases: i) System VAD using the proposed speaker-homogeneous proxy and ii) Oracle VAD using the ground truth speaker-homogeneous speech segments.

We observe that the proposed setup using the VAD system and the speaker-homogeneous proxy approximates the ideal case scenario quite well, reflected in nearly equivalent mAP values for the two cases for most videos. On average, the system's performance attains nearly 91\% of the performance of the ideal case.

\subsubsection{\textbf{Time complexity vs performance}}
In section~\ref{methods:stage1} we proposed to split the video into $p$ partitions and maximize the objective function, $Corr(FD)$ for each partition independently to reduce the time complexity of the system to $O(Ek\frac{N^3}{p^2})$. It assumes that the obtained partitions have enough information to disambiguate the active speakers. For implementation purposes, we partition a video to have at maximum $L$ speech segments in each partition. Using $p=\frac{N}{L}$, we represent the time complexity of the system as $O(EkNL^2)$, making it a linear function of the video's length, $N$ and dependent on the parameter $L$. The length of a partition $L$ signifies the context available to the video split to solve the speech-face assignments, implying that the larger values of $L$ likely provide more disambiguating information and are ideal for the setup. In contrast, selecting a larger $L$ increases the time complexity of the system, thus trading off the time for performance. 

\begin{figure}[]
    \centering
    \includegraphics[width=0.45\textwidth,keepaspectratio]{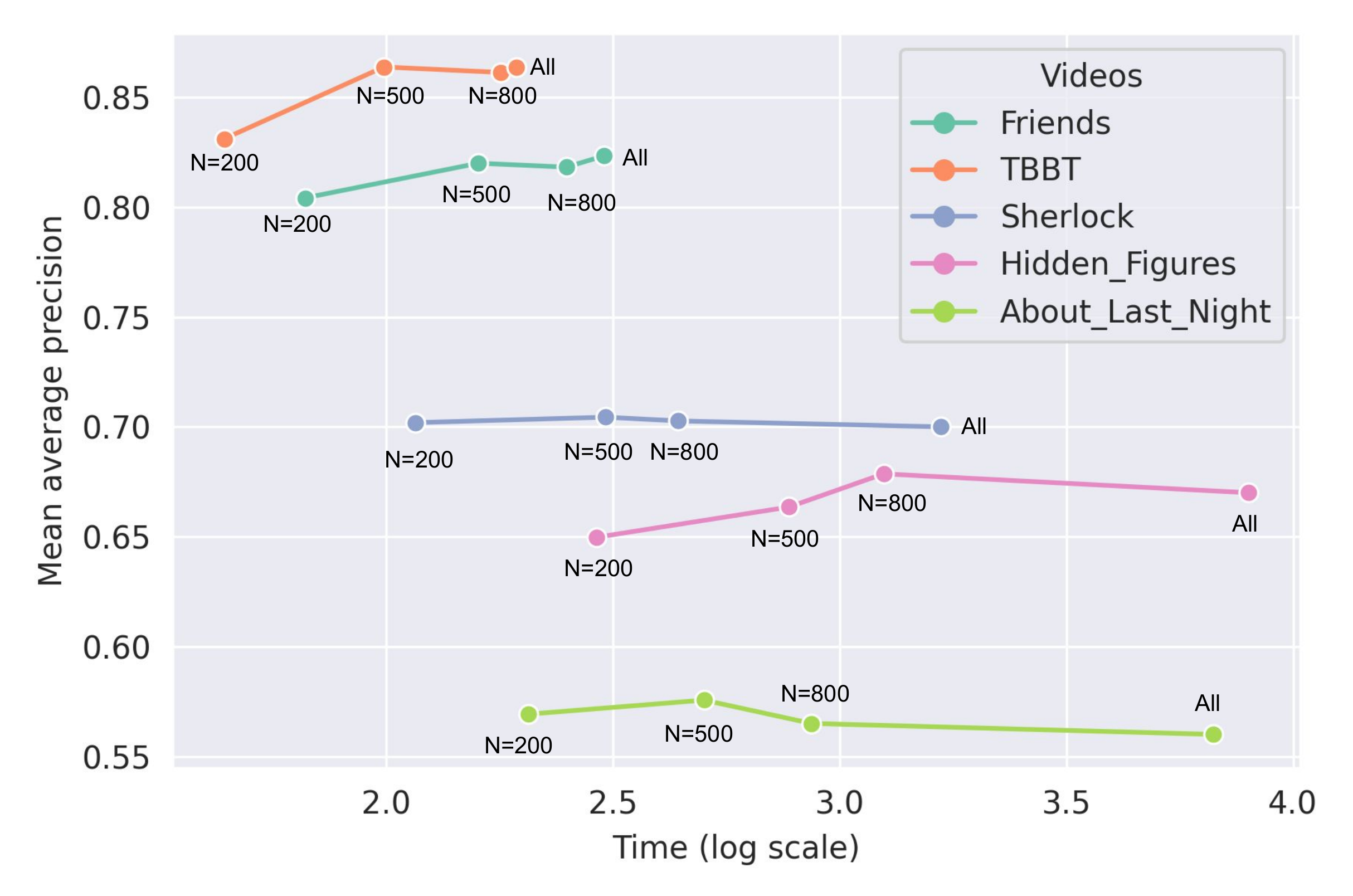}
    \caption{Computational time (in logarithmic scale) for the system against the performance (reported in mAP) for various values of the partition length $L$ for the videos in VPCD.}
    \label{fig:time_complexity}
\end{figure}

In Figure~\ref{fig:time_complexity} we compare the performance of the system (reported in mAP) and the computational time for various values of \mbox{$L=[200, 500, 800]$} and for using total video length at once $(L=N)$. For visualization purposes, we plot the computational time on a logarithmic scale. We observe a significant performance increase for $L=500$ compared with $L=200$ consistently for all the videos in VPCD. Further increasing $L$, we see marginal improvement in the performance along with a significant increase in computational time. Therefore we use $L=500$ for all the videos providing enough context to the video partitions with reasonable computational time. Interestingly we also observe a marginal decrease in performance for the movies, Hidden Figures and About Last Night, and the TV show Sherlock when using the total video length $(L=N)$. This drop in performance is due to a more significant amount of noise in the form of off-screen speakers (Table~\ref{tab:auROC}) compared to other videos, which accumulate while using a larger context. 

\subsubsection{\textbf{Effects of speech and face recognition quality}}
\label{ablation:speech_recognition}
The central idea of this work relies on the fact that speech and face both possess identity information and harness the certitude that the speaker's face and the underlying speech identify the same character. It makes capturing the identity information from speech and face fundamental to the system. Here we study the effect of the quality of captured speech and face identity information on the system's performance. We use two state-of-the-art speech recognition systems to construct different speech identity representations: i) ResNet-34 (default to the system) pretrained on Voxceleb2 dataset displaying 1.2\% equal error rate (EER) for speaker recognition task on Voxceleb1 test set. ii) X-vector embeddings extracted from a TDNN model trained on Voxceleb1+Voxceleb2 training data and displays an EER of 3.2\% on the same Voxceleb1 test set. Note that the lower the EER better is the speaker recognition performance; thus, ResNet-34 performs better for capturing speaker identity information from speech than x-vectors. 

\begin{figure}[]
    \centering
    \includegraphics[width=0.45\textwidth,keepaspectratio]{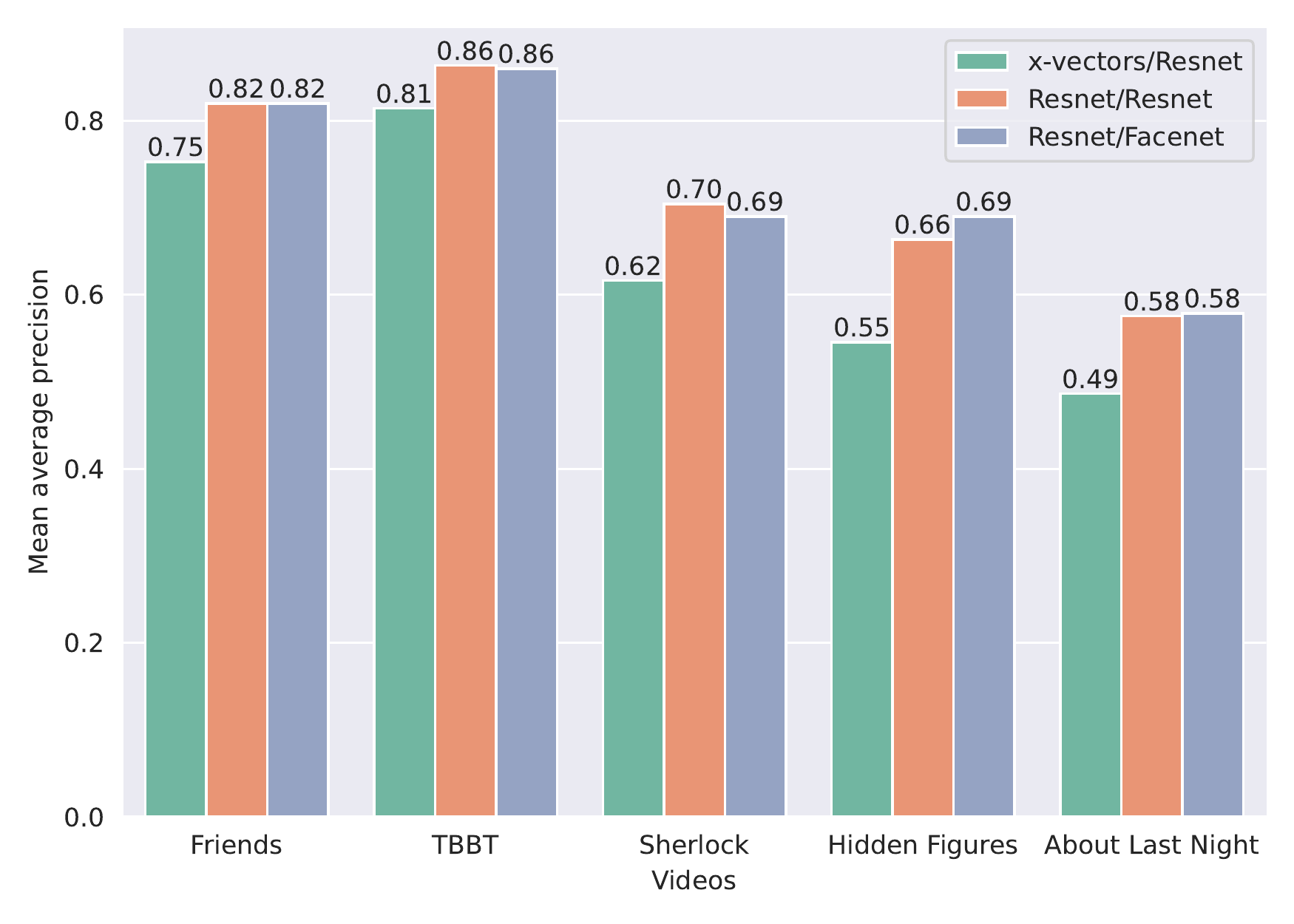}
    \caption{Comparison of active speaker detection performance for systems using different speech/face recognition architectures, on videos from the VPCD dataset. For speech embeddings, Resnet is superior to x-vectors and for face Resnet is superior to the Facenet.}
    \label{fig:xectors}
\end{figure}

In Figure~\ref{fig:xectors}, we show the performance of the active speaker detection system for videos from VPCD, utilizing the two speaker recognition systems. We observe that the performance displayed by the sub-optimal x-vectors embeddings is consistently lower for all the videos than the ResNet-derived embeddings. Using a non-parametric ManWhitneyU test, we noticed that the difference between the performance of the two systems is statistically significant. This implies that the sub-optimal speaker recognition features lead to degraded active speaker recognition performance. It further explains that the low performance for the AVA validation set (reported in Table~\ref{tab:AVA}) is due to the sub-optimal speaker identity embeddings attributed to the noisy speech conditions.

Similarly, to capture the impact of the face recognition quality, we use two state-of-the-art face recognition systems to construct the face identity representations: i) ResNet-50 trained on VGGFace2 (default to the system) and ii) Facenet trained on VGGFace2, compared on IJB-c displaying 0.95 and 0.66 True acceptance rate (TAR) respectively. Higher TAR implies better face recognition performance, pointing out that the Resnet-50 is a superior face recognition system to the Facenet. In Figure~\ref{fig:xectors} we present the active speaker detection performance comparison of systems employing the two different face recognition systems and the same speaker recognition system (ResNet-34). We observe that the difference in performance is marginal and shows no statistical significance. It suggests that the system is more tolerable to degradation in face recognition performance. 

\subsection{Discussion}
The proposed framework relies on the assumption that the video under consideration has enough information to disambiguate the characters in the video. The presence of scenes with just the speaking character in the frames is one of the simple examples of such character disambiguating information, as it provides an instant speech-face association. The videos intending to capture the speaking characters will likely have relevant information to derive the speech-face association based on the co-occurrence of the speech and faces through the video. On the contrary, one of the adverse limiting cases for the system is a video with a fixed set of faces appearing in the frames at all times. A simple example can be a panel discussion shot with a static camera, capturing all the panelists at all times. In such a video, the system will have no way to correctly associate the identity from the speech with one of the faces on the screen. Columbia dataset~\cite{chakravarty2016cross}: a video of a panel discussion is very similar to such a scenario. Even though the video is  85 min long, the annotations are available for 35 min duration, comprising six speakers. 

\begin{figure}[]
    \centering
    \includegraphics[width=0.45\textwidth,keepaspectratio]{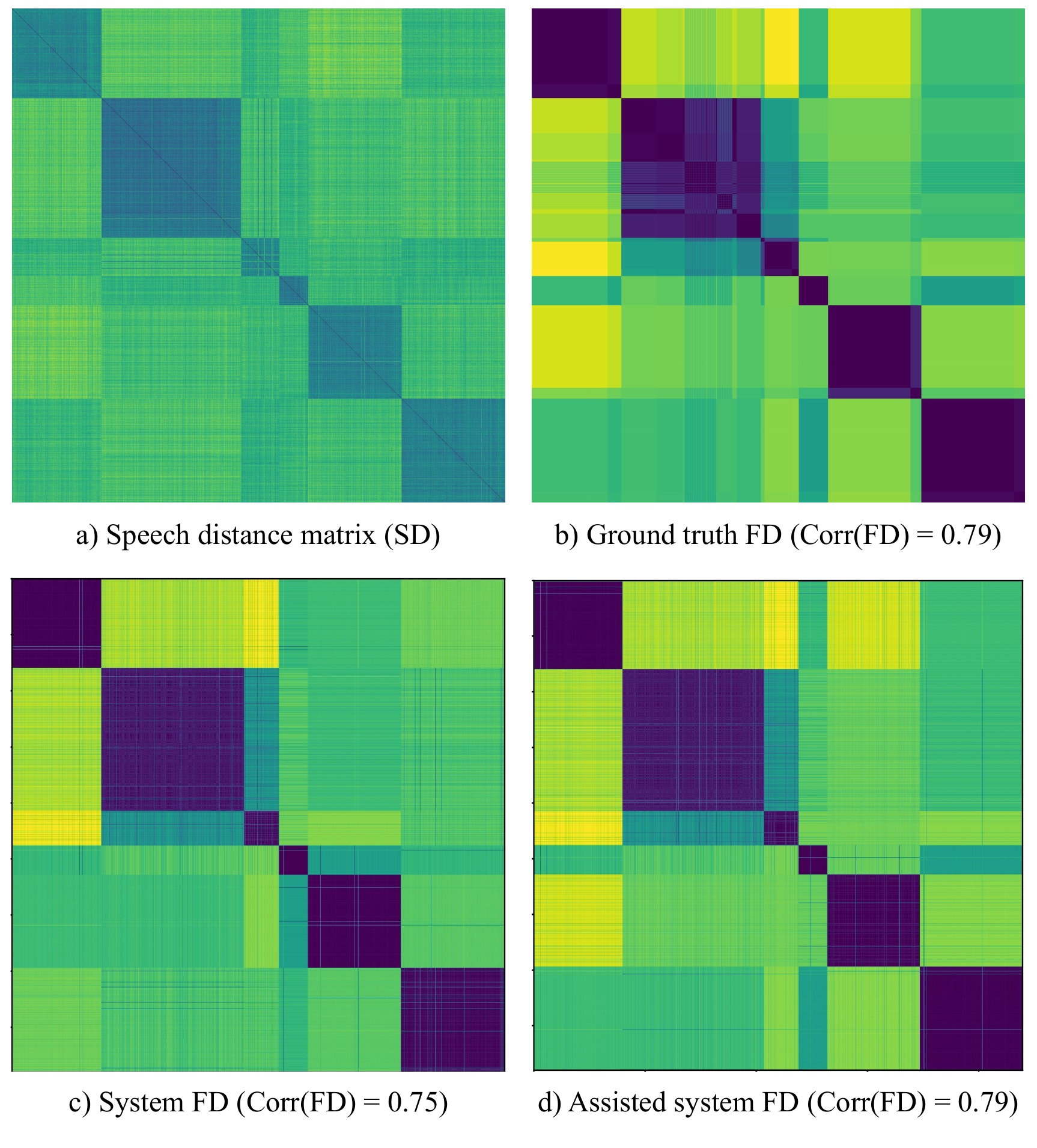}
    \caption{Distance matrices for Columbia dataset  with $Corr(FD)$. a) Speech distance matrix $SD$. b) $FD$ with ground truth active speaker faces. c) $FD$ post stage2 for proposed system. d) $FD$ post stage 2 for the system assisted with 15\% ground truth active speaker faces.}
    \label{fig:columbia_distances}
\end{figure}

We evaluate the performance of the proposed system on the Columbia dataset and report the speaker-wise weighted F1-score, a widely used metric for evaluating this dataset, in Table~\ref{tab:columbia}, along with performance reported by various other methods. We observe that the F1 score is pretty low for three speakers (Lieb, Long, and Sick), while for the other three, it is at par with other systems. We further investigated and observed that the face distance matrix $FD$ obtained using the predicted active speaker faces attains a high value of the objective function $Corr(FD)$. The speech ($SD$) and the face distance ($FD$) matrices are shown in Figure~\ref{fig:columbia_distances}a \& c. The obtained $Corr(FD)$ is close to the one displayed by the ground truth active speaker faces, shown in Figure~\ref{fig:columbia_distances}b, suggesting the intended working for the proposed system.

\begin{table}[]
\centering
\caption{Comparison of the speaker-wise weighted F1 scores for all the speakers in Columbia dataset.}
\label{tab:columbia}
\resizebox{0.45\textwidth}{!}{%
\begin{tabular}{@{}ccccccc|c@{}}
\toprule
Methods             &Abbas  & Bell & Boll  & Lieb & Long & Sick & Avg   \\ \midrule
Chakravarty et al.~\cite{chakravarty2016cross}  &-      & 82.9 & 65.8  & 73.6 & 86.9 & 81.8 & 78.2  \\
Shahid et al~\cite{shahid_columbia}          &-      & 89.2 & 88.8  & 85.8 & 81.4 & 86   & 86.2  \\
Chung et al.~\cite{chung2016out}             &-      & 93.7 & 83.4  & 86.8 & 97.7 & 86.1 & 89.5  \\
Afouras et al.~\cite{afouras2020self}     &-      & 92.6 & 82.4  & 88.7 & 94.4 & 95.9 & 90.8  \\
S-VVAD~\cite{svvad}           &-      & 92.4 & 97.2  & 92.3 & 95.5 & 92.5 & 94    \\
RealVAD~\cite{realvad}           &-      & 92   & 98.9  & 94.1 & 89.1 & 92.8 & 93.4  \\
TalkNet ~\cite{talknet}           &-      & 97.1 & 90.0    & 99.1 & 96.6 & 98.1 & 96.2  \\ \midrule
Proposed            &97.0  & 95.0 & 91.0 & 65.3  & 0.1 & 34.6  & 65.0 \\
Proposed (assisted) &96.7  & 95.3 & 90.5  & 98.2 & 93.2 & 96.1 & 95.0  \\ \bottomrule
\end{tabular}%
}
\end{table}

\begin{figure*}[]
    \centering
    \includegraphics[width=\textwidth,keepaspectratio]{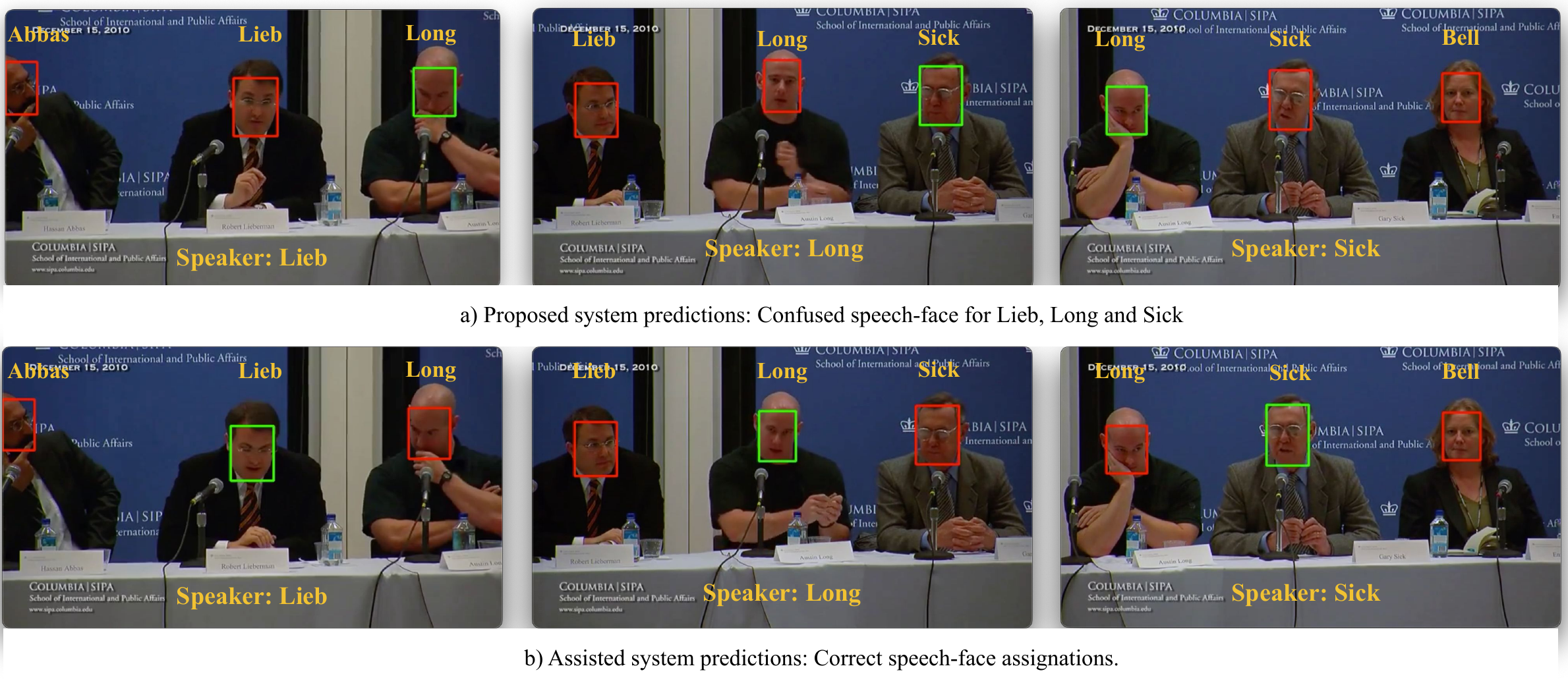}
    \caption{Sample frames from Columbia dataset with manually marked identities of the faces and the name of the current speaker. a) The frames shows the incorrect speech-face associations by the proposed system. b) The frames shows the corrected speech-face associations when assisted with 15\% ground truth faces}
    \label{fig:columbia_samples}
\end{figure*}

On visual inspection of the predicted active speaker faces, we observed that the system incorrectly associates Long's speech with Sick's face and Lieb's with Long's face. The system selects Sick's face as an active speaker whenever Long speaks and similarly for the other two speakers. Relevant frames from the video with the system's predictions (green box: active speaker) and manually marked speakers' identities are shown in Figure~\ref{fig:columbia_samples}a. We further observed that Sick always appears with Long throughout the video, providing the system with insufficient information to resolve the speech-face association. Similarly, Lieb always appears with Long, limiting the system to solve the discrepancy.

To close the loop on the system's performance on the Columbia dataset, we simulate a setup by manually forcing disambiguating information into the video, which assists the system in solving the speech-face association. We randomly select 15\% of the speech segments of each speaker and initialize the corresponding active speaker faces using the ground truth active speaker faces. While maximizing the objective function $Corr(FD)$ (using Algorithm~\ref{algo:optimzation} in stage1), we forbid updating the active speaker face assignment for the selected speech segments. It replicates the trivial case of a scene with just the speaker's face visible in frames, thus synthetically injecting information to resolve speech-face associations. We show the face distance matrix ($FD$) for the obtained active speaker faces in Figure~\ref{fig:columbia_distances}d and note that the objective function $Corr(FD)$ is nearly equivalent to the one shown by the ground truth active speaker faces (Figure~\ref{fig:columbia_distances}b). We report speaker-wise F1-score in Table~\ref{tab:columbia} and observe that the assisted system performs similar to state-of-the-art methods for all the speakers. In Figure~\ref{fig:columbia_samples}b we show the frames with corrected speech-face associations. Video clips highlighting the system's initial incorrect active speaker predictions and the later obtained corrected predictions using the assistance from the ground truth are present in supplementary material. We point out that this easy way of incorporating external information can be extended to integrate complementary information from visual-activity-based systems to generate a comprehensive solution to active speaker detection.

%% file: conclusion.tex
\section{Summary and Conclusion}
\label{sec:conclusion}


We presented a cross-modal unsupervised framework for active speaker detection in videos, harnessing the character identity information in the speech and the faces of the speakers. We evaluated the proposed system on three benchmark datasets, consisting of various videos from entertainment media (TV shows and movies) and broadcast media (panel discussion), and showed competitive performance compared to state-of-the-art fully supervised methods. The framework utilizes off-the-shelf speech and face identity representation systems and shares their limitations when speech accompanies noise and music. 

The framework provides an easy way to inject external information, enabling it to collaborate with other methods. It can incorporate high-confidence predictions of other state-of-the-art methods and even inputs from expert humans in the loop, enabling a provision to provide supervision at various degrees. One of the  future directions can be to include the complementary set of information from visual-activity-based methods providing more disambiguating information to the system and thus further enhance the performance.